\documentstyle[prl,aps,preprint]{revtex}
\begin{document}
\tighten
\title{FORM FACTOR DESCRIPTION OF THE NON-COLLINEAR 
COMPTON SCATTERING TENSOR} 

\author{Wei Lu}

\bigskip

\bigskip

\address{
Centre de Physique Th\'eorique\footnote{
Unit\'e propre 14 du Centre National de la Recherche Scientifique.},   
Ecole Polytechnique\\ 
91128 Palaiseau  Cedex, 
France \\ 
{~}}

\date{July 1997}

\maketitle

\begin{abstract}

We present a parameterization of the non-collinear 
(virtual) Compton scattering tensor in terms of form factors, 
in which  the Lorentz tensor associated with each form factor 
possesses manifest electromagnetic gauge invariance. 
The  main finding is that 
in a well-defined form factor expansion of the scattering tensor,
the form factors are  either symmetric or antisymmetric under 
the exchange of two Mandelstam variables,   $s$ and $u$. 
Our decomposition can be used to    organize
 complicated  higher-order and higher-twist 
contributions in the  study of the 
virtual Compton scattering off the proton.  
Such procedures  are  illustrated 
by use of the  virtual Compton scattering off the lepton.  
In passing, we note the general symmetry constraints 
on Ji's  off-forward  parton distributions and Radyushkin's 
double  distributions. 

\end{abstract}

\pacs{PACS Numbers: 9.65,+i, 8.38.Aw, 9.85.Qk}

\section{introduction} 

Recently, there  is \cite{ji1,ra1,hyde,chen,pire}  
much revived interest in the  virtual 
Compton scattering (VCS). 
By VCS,  people usually mean the the  scattering  of a 
virtual photon into a real photon off a proton target
$$ \gamma^\ast (q) + N(P,S) \to \gamma^\ast(q^\prime) 
+N(P^\prime, S^\prime) \ . $$ 
As usual, three Mandelstam variables are defined for this process:
 $s \equiv (q+P)^2, $ 
$  t \equiv  (q-q^\prime)^2 ,$ 
$ u \equiv(P-q^\prime )^2$. 
Due to the momentum conservation 
$
P + q = P^\prime + q^\prime, 
$
there  is  the  following constraint: 
\begin{equation}
s+t +u =q^2  + q^{\prime 2} + 2m^2,       
\end{equation} 
where $m$ is the proton mass. 

The object of study is  the following scattering tensor 
\begin{equation}
T^{\mu\nu}(q, P,S; q^\prime,  P^\prime,S^\prime)
=
i\int  d^4 \xi 
e^{iq^\prime\cdot \xi}
\langle P^\prime,S^\prime| {\rm  T}[J^\mu (\xi)
J^\nu (0)]|P,S\rangle \ ,   
\end{equation}
where $J$ is the quark electromagnetic current in the proton and T stands 
for the time-ordering of the operators. At present, 
 much of interest is focused on  the deeply VCS (DVCS),  which is 
a very special kinematic region of the generic VCS.  It has been claimed that  
the dominant mechanism in the DVCS is the VCS off a massless quark
\cite{ji1}. Correspondingly,  two different approaches to the 
DVCS  tensor have been developed:  
 the Feynman diagram expansion \cite{ji1,ra1} and  operator
product expansion  (OPE) \cite{wana,chen}. 

  A careful reader might be aware of such a fact: At  the 
leading  twist expansion  of the  DVCS tensor, 
both in the Feynman diagram expansion  and in OPE approach,  the resultant 
expressions  do not possess   manifest electromagnetic gauge invariance.
The purpose of this paper  is to remedy the case by 
presenting a full  form factor parameterization of the 
non-collinear  Compton scattering tensor. 
With the help of our decomposition of the scattering tensor, 
one can safely ignore the higher-twist terms at leading-twist expansion
 and  recover the electromagnetic 
gauge invariance by brute force. Hopefully, our form factor 
description can be used to organize complicated 
 calculations as one goes  beyond 
leading twist and/or  leading order. 
 
 We confess that we are not the very first to attempt to 
develop a form factor parameterization of the   VCS
 tensor. As early as in 1960s,  Berg and Lindner 
\cite{lindner} ever reported a 
parameterization of the VCS tensor in terms of form factors. 
The virtue, also an implicit assumption, in their 
decomposition  is that  the scattering tensor can be 
put into  a form of direct products of the 
Lorentz tensors and Dirac bilinears, i.e., the  Lorentz index of the 
VCS tensor is $not$ carried by the gamma matrices.
In fact,  all the leading twist expansions of the DVCS tensor 
so far  assume such a  factorized form.  A drawback of the   
Berg-Lindner decomposition is that they  employed a lot of momentum 
combinations  which have no specific crossing and time reversal 
transformation properties. As a consequence, the form factors  they defined 
possess no specific symmetry properties under crossing and time reversal 
transformations.  Moreover,  their decomposition  lacks a term  
associated  with the 
Lorentz structure $\epsilon^{\mu\nu\alpha\beta}q_\alpha q^\prime_\beta$, 
which has been shown by recent researches 
 to be a carrier of leading-twist contributions.

 It should be stressed that there is no unique decomposition of 
the Compton  tensor. A few years ago, Guichon, Liu and Thomas 
\cite{pierre}  worked out a  general decomposition of the 
VCS tensor, which  contains no   explicit proton spinors.  
Their decomposition is nice for the discussion of the  generalized 
proton polarizabilities, as has done in Ref. \cite{pierre}.  Recently, 
the decomposition of the VCS tensor of such type has been refined 
by Drechsel et al. \cite{Drechsel}
 within more extensive contexts. However,  
 a  decomposition of the  VCS  tensor without explicit 
Dirac bilinear structures   is of  very limited use for the  present 
Feynman diagram expansion and OPE analysis of the DVCS tensor.

 Hence, it is desirable to reconstruct a  parameterization 
of the Compton scattering tensor in terms of 
form factors with explicit Dirac structures, which 
constructs the subject  of this paper. 
To make our arguments more transparent, we will  first consider
the Lorentz decomposition for  the double  VCS off a lepton, 
then transplant  our results 
 onto the proton case. [By double VCS we mean that 
both the initial- and final-state photons are virtual. Correspondingly, 
we will refer the usual VCS to as the single VCS in
distinction.]  The reason  for adopting such a strategy 
is that in quantum electrodynamics, it is  more convenient to 
discuss the chiral properties  of the Dirac bilinears. 
At the later stage, we will reduce our results for the double VCS 
 to the real Compton scattering (RCS) as well as the single VCS. 
 Such a procedure  will greatly facilitate the discussion of 
the symmetry properties of the single  VCS form factors.

 The decomposition of the Compton  tensor is  essentially  subject to 
 the symmetries  that it observes, so we  begin with a 
brief  discussion of the symmetry properties of the 
Compton scattering. First,  the  current  conservation  requires that 
\begin{equation}
q_\mu T^{\mu\nu}(q, P,S; q^\prime,  P^\prime,S^\prime) 
= 
q^\prime_\nu T^{\mu\nu}(q, P,S; q^\prime,  P^\prime,S^\prime) 
=0\ .   \label{gauge} 
\end{equation}
Second,  parity conservation tells us 
\begin{equation}
 T^{\mu\nu}(q, P,S; q^\prime,  P^\prime,S^\prime) 
= 
 T_{\mu\nu}( \tilde q,\tilde P, - \tilde S; \tilde q^\prime,  
\tilde P^\prime, - \tilde S^\prime) \ , \label{p}       
\end{equation}
where $\tilde q^\mu \equiv q_\mu$, and so on. 
Third,   time reversal invariance demands 
\begin{equation}
 T^{\mu\nu}(q, P,S; q^\prime,  P^\prime, S^\prime) 
= 
 T_{\nu\mu}(\tilde  q^\prime, \tilde  P^\prime,\tilde S^\prime; 
\tilde q, \tilde P,\tilde S )\  .  \label{t}   
\end{equation}
Fourthly,  there is a  crossing symmetry for the Compton scattering, namely, 
\begin{equation}
T^{\mu\nu}(q, P,S; q^\prime,  P^\prime, S^\prime) 
= T^{\nu\mu}(-q^\prime, P,S; -q,  P^\prime, S^\prime)\ . \label{cr} 
\end{equation} 

By combining (\ref{p}) with (\ref{t}), we have 
 \begin{equation}
T^{\mu\nu}(q, P,S; q^\prime,  P^\prime, S^\prime) 
= T^{\nu\mu}(q^\prime, P^\prime, -S^\prime ;  q, P,-S)\ .\label{pt} 
\end{equation} 
That is to say,  the adjoint parity-time-reversal transformation
amounts to $\mu \leftrightarrow \nu$, 
$q \leftrightarrow q^\prime $, 
$P \leftrightarrow P^\prime $, 
$S\to -S^\prime$ and  $S^\prime \to -S$. 
Furthermore, combining (\ref{cr})  with (\ref{pt}) yields 
\begin{equation}
T^{\mu\nu}(q, P,S; q^\prime,  P^\prime, S^\prime) 
= T^{\mu\nu}(-q,P^\prime,- S^\prime  ; -q^\prime, P,-S  )\ . \label{crpt} 
\end{equation}

In fact, the Compton scattering  respects  more symmetries
than summarized above. For example, it  is    
subject to  the momentum  and angular momentum conservations. 
In the   case of  collinear scattering, the angular 
momentum conservation exerts further constraints on
the Compton scattering. To show this, we  digress 
to the helicity amplitude description of the Compton scattering. 

In the expansion of the Compton scattering amplitude, a
fundamental question that must be answered in advance is that 
how many independent state vectors there are in a complete basis. 
This can be most naturally done by counting independent 
helicity amplitudes.  Here we stress that each independent  helicity 
amplitude corresponds to one $observable$ independent form factor in the 
Compton scattering tensor, while there is no simple  one-to-one 
correspondence between helicity amplitudes and  form factors.

  Let us  consider  the most general non-collinear
double VCS off a massive lepton ($l$). Because the massive lepton  and 
the virtual photon have 2 and 3 helicity states respectively, 
there are $2\times 3 \times 2 \times 3=36$ helicity amplitudes. 
By parity conservation, only  half  of the  these helicity amplitudes
are independent. In the non-collinear case,  
the other symmetries cannot  further  reduce the 
number of independent helicity amplitudes.
Similarly, there are 12 and 8  independent  helicity 
amplitudes for the non-collinear  single   
VCS and RCS off the massive lepton. 

In  the collinear scattering limits, however, 
time reversal invariance and angular momentum conservation 
impose further  constrains on the Compton scattering. 
We denote  a generic Compton scattering  helicity amplitude 
$A(\lambda_q, \lambda_l; \lambda_{q^\prime}, \lambda_{l^\prime})$, 
where $\lambda$'s are the helicities of the corresponding particles. 
By time reversal  invariance, there  is 
\begin{equation} 
A(\lambda_q, \lambda_l; \lambda_{q^\prime}, \lambda_{l^\prime})
=
  A( \lambda_{q^\prime}, \lambda_{l^\prime}; \lambda_q, \lambda_l)\ . 
\end{equation}
To discuss the constraints from angular momentum conservation, 
we need to distinguish  two  collinear limits: 
\begin{equation} 
\lambda_q -\lambda_l=\pm (\lambda_{q^\prime} -\lambda_{l^\prime}) \ . 
\end{equation} 
where $\pm$ corresponds to the forward  and backward
collinear scattering, respectively. As a result,  
only a small fraction of the Compton helicity amplitudes survive 
in the collinear limits. We summarize those 
surviving (independent) helicity amplitudes in  Table 1.

 The above helicity anplitude  analysis implies a thorny fact: As one 
approaches the collinear limits, there is significant degeneracy 
in the form factor parameterization of the Compton scattering 
tensor.  Here we emphasize  that one must  avoid 
the over-degeneracy  of  the form factor description in the 
collinear cases as much as possible.  As  will be  clarified, 
some  form factor parameterizations of the Compton 
scattering tensor, albeit applicable in the non-collinear cases, 
might become ill-defined in the collinear scattering limits. 

 Now we investigate the general structure of the 
non-collinear Compton tensor.  As stated before, the Berg-Lindner 
decomposition assumes  the following form 
 \begin{equation}
T^{\mu\nu}(q, P,S; q^\prime,  P^\prime, S^\prime) 
= \sum_{i}    \bar U(P^\prime, S^\prime) 
\Gamma_i U(P,S)t^{\mu \nu}_i F_i \ , \label{6}
\end{equation}
where  $\Gamma_i$s are gamma matrices 
(saturated  with particle momenta if  carrying Lorentz indices),
$t^{\mu \nu}_i$s   Lorentz (pseudo)-tensors constructed
from  the relevant particle momenta (the metric tensor and the 
Levi-Civita tensor may be involved), and 
$F_i$s   Lorentz invariant form-factor like objects. 
As a matter of fact, all of the  recent research results 
about  the DVCS tensor can be tailored into the form of Eq. (\ref{6}).

Now we justify Eq. (\ref{6})   for the non-collinear 
Compton scattering. In principle, one can
  assume a  decomposition for  the
Compton tensor  in which  there is no 
explicit  Dirac bilinears. Then, the Lorentz indices 
of the scattering tensor  can be carried by  
the metric tensor, the Levi-Civita tensor,  the particle 
momenta and    lepton   spin four-vectors. 
We will not write down any such  decompositions. 
Rather, we note that the lepton    spin
four-vector can carry the free Lorentz index now.
The  spin four-vector $S$ of a lepton of momentum $P$ 
 is subject to $S\cdot P=0$,
so it can be expressed in terms of three $non$-$collinear$ particle 
momenta. For  the non-collinear  scattering, one can write down, 
  say 
 \begin{equation}
S^{\mu}=(S\cdot K_1)  P^{\prime \mu} +(S\cdot K_2)
 q^\mu + (S\cdot K_3)q^{\prime \mu}, \label{7}   
\end{equation}  
where $K_1,$ $K_2$ and $K_3$ are  three  momentum combinations 
whose expressions we do not need. 
As a consequence,  one can eliminate  the lepton  spin four-vectors 
from  the building blocks that carry the free Lorentz indices and 
lump $S\cdot K_j$ into  the form factors. At this stage, if one
displays the Dirac bilinears,  the  decomposition  of the 
Compton  scattering tensor  assumes the structure  of Eq. (\ref{6}). 

 From the above justification,   we see 
that some subtleties will  arise as one goes to the collinear limits
of the Compton scattering. 
That is,  Eq. (\ref{6}) is inapplicable to the 
discussion of the transverse proton spin  dependence of the
Compton scattering  amplitude
in the collinear limits. Fortunately, the collinear scattering are 
only  very special kinematic limits of the VCS. So it is  still 
desirable to develop a  form factor parameterization of the 
non-collinear Compton scattering  with the general 
 structure  of Eq. (\ref{6}).

  The  symmetries impose further   constraints 
on the decomposition of Eq. (\ref{6}).   For a generic  
VCS,  its form factors  depend on 
4  independent kinematical variables. 
Though there is  much interest in the small-$|t|$ limit 
behavior \cite{ji1,pire} of the single VCS off the proton,  
we insist in  choosing  $s$, $u$, $q^2$ 
and $q^{\prime 2}$ ($t$ being an auxiliary quantity) 
 as four independent kinematical  variables
for the form factors. The reason for doing so is that 
the crossing transformation of the Compton scattering 
 essentially relates its $s$- to $u$-channel contributions
or vice versa. Under the crossing transformation, 
$ s  \leftrightarrow u$ and  
$q^2 \leftrightarrow q^{\prime 2}$.  Further, under the time 
reversal (or the  adjoint parity-time-reversal) transformation, 
there is  $q^2 \leftrightarrow q^{\prime 2}$.  
So,  it is a natural choice  for us 
to define the form factors in such a way that all of them possess 
specific symmetry properties under the crossing and time-reversal 
transformations. To this end, we demand
that the  Lorentz tensor and Dirac bilinear 
 associated with each form factor are either symmetric or 
antisymmetric under  the crossing and time-reversal 
transformations.  At this stage, we recognize that 
 it is more useful to talk 
about the adjoint crossing-(parity)-time-reversal transformation 
properties of the single VCS form factors, for which people usually 
take into account the on-shell condition $q^{\prime 2}=0$ of the 
final photon  in practical calculations.

  Now we  set about the construction of the 
form factor description of the non-collinear Compton scattering tensor. 
As claimed earlier, we begin with  the  double VCS off a   lepton. 
We first consider the  case of a  massless lepton.  
Then, its helicity and chirality coincide with each other. 
For  the massless lepton,   the chiral symmetry holds
exactly, so  the lepton   helicity  is conserved  in  the 
Compton scattering.  As a  consequence, there 
are  only 9 independent helicity amplitudes and accordingly 
9 complex  form factors for the  non-collinear double
VCS off the massless lepton.  All the spinor bilinears must 
be chiral-even, so only the vector and axial-vector Dirac 
structures, $\gamma_\alpha  $ and $ \gamma_\alpha\gamma_5,$ 
 get into work. 
To saturate  the Lorentz indices carried by the  Dirac matrices, 
we choose  $q+q^\prime$, $P$ and $P^\prime$ as  3 independent momenta.     
Obviously,  there are only  two nontrivial, independent
Dirac structures: $ \bar U(P^\prime,S^\prime)(\rlap/q+\rlap/q^\prime)   
U(P,S)$ 
and $ \bar U(P^\prime,S^\prime)(\rlap/q+\rlap/q^\prime)  \gamma_5  U(P,S)$. 

 Now we construct proper gauge-invariant Lorentz tensors 
to match  $ \bar U(P^\prime,S^\prime)(\rlap/q+\rlap/q^\prime)  U(P,S)$.
In doing so, we keep it in mind to render our choices of 
independent Lorentz tensors possess specific crossing and 
parity-time-reversal transformation properties. 
Using the metric tensor, we can write down 
$ -(q\cdot q^\prime)g^{\mu\nu}+q^{\prime\mu} q^\nu$. As 
index $\mu$ is  carried by particle momenta,  we can write 
down only two independent momentum combinations because of the 
momentum conservation and gauge invariance. Similarly for index $\nu$. 
So, we have 4 more  independent tensors without invoking the 
metric tensor. Our choices are  
 $ A^\mu B^\nu$, $ A^\mu_1 B^\nu + A^\mu B^\nu_1$, 
$ A^\mu_1 B^\nu - A^\mu B^\nu_1$, 
and  $ A^\mu_1 B^\nu_1,$ where 
\begin{eqnarray} 
  A^\mu&=& (q^{\prime \mu} - \frac{q\cdot q^\prime}{P\cdot q } P^{ \mu})
+(q^{\prime \mu} - \frac{q\cdot q^\prime}{P^\prime\cdot q } P^{ \prime \mu})
\ ,\\ 
 A^\mu_1&=&q^\mu -\frac{q^2}{q\cdot q^\prime}q^{\prime\mu}\ ,\\  
B^\nu&=&(q^{ \nu} -\frac
{q\cdot q^\prime }{P^\prime \cdot q^\prime  }P^{ \prime \nu})
+(q^{ \nu} -\frac
{q\cdot q^\prime }{P \cdot q^\prime  }P^{  \nu})
 \ , \\  
B^\nu_1&=&q^{\prime \nu}  -\frac{q^{\prime 2} }{q\cdot q^\prime}q^{\nu}\ .  
 \end{eqnarray} 
By construction,  $ A^\mu B^\nu$, $ A^\mu_1 B^\nu + A^\mu B^\nu_1$, 
$ A^\mu_1 B^\nu - A^\mu B^\nu_1$, 
and  $ A^\mu_1 B^\nu_1 $ have specific symmetry properties 
under the crossing and time-reversal transformations.

 To  match  $ \bar U(P^\prime,S^\prime)(\rlap/q +\rlap/q^\prime)
\gamma_5  U(P,S)$, 
we need to invoke one Levi-Civita tensor.  If    the Levi-Civita
tensor is demanded to carry two free Lorentz indices, we have 
$\epsilon^{\mu\nu\alpha \beta }q_\alpha  q^\prime_\beta$  by gauge
invariance.   As one of  the Lorentz indices is carried by the
particle  momentum,  at our disposal  are 
$A^\mu D^\nu +B^\nu C^\mu,$ $A^\mu D^\nu -B^\nu C^\mu,$
 $A^\mu_1 D^\nu+B^\nu_1 C^\mu$, $A^\mu_1 D^\nu-B^\nu_1 C^\mu$,   where 
\begin{eqnarray} 
  C^\mu&=&\epsilon^{\mu \alpha \beta \gamma }q_\alpha 
 P_\beta   P ^\prime_\gamma\ ,\\ 
 D^\nu&=&\epsilon^{\nu \alpha \beta \gamma }q^\prime_\alpha 
 P_\beta   P ^\prime_\gamma\  . 
 \end{eqnarray} 
Again, 
$A^\mu D^\nu +B^\nu C^\mu,$ $A^\mu D^\nu -B^\nu C^\mu,$
 $A^\mu_1 D^\nu+B^\nu_1 C^\mu$, $A^\mu_1 D^\nu-B^\nu_1 C^\mu$ 
have specific symmetry properties under the 
crossing and time-reversal transformations.

By definition,  $A^\mu D^\nu,$  $A^\mu_1 D^\nu,$ $B^\nu C^\mu$,
 and  $B^\nu_1 C^\mu$ are independent of  each other.
On the other hand, the  
antisymmetric property  of $\epsilon^{\mu\nu\alpha \beta }
q_\alpha  q^\prime_\beta$  tells  us it  that may have  6 non-vanishing 
components.   Due to  the  current conservation conditions, Eq. 
 (\ref{gauge}), 
only 4 of them are independent.  Therefore, $\epsilon^{\mu\nu\alpha \beta }
q_\alpha  q^\prime_\beta$   can be expanded in terms of 
$A^\mu D^\nu,$  $A^\mu_1 D^\nu, $ $ C^\mu B^\nu$, and  $ C^\mu B^\nu_1$.
In fact, one  can directly  construct the 
following identity: 
\begin{eqnarray} 
\epsilon^{\mu\nu\alpha \beta }
q_\alpha  q^\prime_\beta &= & 
\frac{( P\cdot A_1 -P^\prime\cdot A_1 ) A^\mu D^\nu 
-(P\cdot A -P^\prime\cdot A ) A^\mu_1 D^\nu}  
{(P\cdot A) (P^\prime \cdot A_1)- 
(P^\prime \cdot A)  (P \cdot A_1) } \nonumber\\ 
& & 
+ \frac{ - ( P\cdot B_1 -P^\prime\cdot B_1 ) C^\mu B^\nu 
+(P\cdot B  -P^\prime\cdot B  ) C^\mu B^\nu_1}  
{(P\cdot B)  (P^\prime \cdot B_1)- 
(P^\prime \cdot B)  (P \cdot B_1)}\ .  
\end{eqnarray} 
Notice that this identity holds only for the non-forward
Compton scattering. 
This is where the subtleties arise in choosing four Lorentz 
pseudo-tensors to match  $ \bar U(P^\prime,S^\prime)(\rlap/q +\rlap/q^\prime)
\gamma_5  U(P,S)$.  If one  selects $A^\mu D^\nu$, $ B^\nu C^\mu,$ 
 $A^\mu_1 D^\nu,$   and  $ C^\mu B^\nu_1$, all of them will drop out 
in the collinear limits.  As we stressed, we need to avoid 
the  over-degeneracy in  the collinear limits as much as possible. 
On the other hand, all of recent studies  indicate
that the $\epsilon^{\mu\nu\alpha \beta }q_\alpha  q^\prime_\beta$
term incorporates the leading twist contributions. 
 Hence,  we  choose  $\epsilon^{\mu\nu\alpha \beta }
q_\alpha  q^\prime_\beta$, $A^\mu D^\nu+ B^\nu C^\mu,$ 
 $A^\mu_1 D^\nu+ C^\mu B^\nu_1 ,$   and  
$A^\mu_1 D^\nu- C^\mu B^\nu_1 $  as 4 independent 
Lorentz pseudo-tensors to match with  $ \bar U(P^\prime,S^\prime)
(\rlap/q +\rlap/q^\prime)\gamma_5  U(P,S).$ [In the Berg-Lindner 
decomposition,  there is no 
$\epsilon^{\mu\nu\alpha \beta }q_\alpha  q^\prime_\beta$
term.]
Notice that $\epsilon^{\mu\nu\alpha \beta }q_\alpha  q^\prime_\beta$
survives the collinear limits. 
 
Thus, we have identified 9 independent structures for the 
non-collinear  double VCS off the massless lepton.

Now we take into the lepton mass effects. Then,  the 
helicity of a massive lepton is no longer in coincidence with 
its chirality.  The chiral-even lepton state is roughly 
in the  helicity-$+\frac{1}{2}$ state, with a helicity-$-\frac{1}{2}$ 
contamination of 
${\cal O}(m/Q)$, where $m$ and $Q$ are the lepton mass and a high
energy interaction scale, respectively. The inclusion of the 
lepton mass effect  will generate 
9  more  structures, which flip the lepton helicity. Writing down the 
Dirac bilinears is essentially an expansion according to 
the   chiral structure.
Although the expansion according to the lepton helicity is not 
coincident with that according to the   chiral structure, the 
number of independent terms in any complementary expansion 
should be equal.  Hence, the lepton mass will generate 9 more 
independent chiral-odd structures in the general decomposition. 
 
 In  constructing  independent chiral-odd Dirac bilinears, we have 
$1$, $\gamma_5$ and $\sigma^{\alpha\beta}$ at our disposal. 
We first consider  the pseudo-scalar 
Dirac structure $ \bar U(P^\prime,S^\prime)\gamma_5  U(P,S)$.  
Remind that there should be at least  two spin-dependent form factors 
in the collinear scattering limit, because 4 independent helicity 
amplitudes survive in the collinear limits of double VCS. 
To avoid possible over-degeneracy in the collinear limits, 
 we  select again   
 $\epsilon^{\mu\nu\alpha \beta }
q_\alpha  q^\prime_\beta$, $A^\mu D^\nu+ B^\nu C^\mu,$ 
 $A^\mu_1 D^\nu+ C^\mu B^\nu_1 ,$   and  
$A^\mu_1 D^\nu- C^\mu B^\nu_1 $  as 4 independent 
Lorentz pseudo-tensors to match with 
$ \bar U(P^\prime,S^\prime)\gamma_5  U(P,S)$.

Now we consider the tensor Dirac structures. By choosing 
$q+q^\prime$, $P$ and $P^\prime$  as 3  independent particle 
momenta, we have 
$ \bar U(P^\prime,S^\prime)\sigma^{\alpha\beta}P_\alpha 
P^\prime_\beta  U(P,S),$
$\bar U(P^\prime,S^\prime)\sigma^{\alpha\beta}(q +q^\prime) 
_\alpha P^\prime_\beta  U(P,S),$  
$\bar U(P^\prime,S^\prime)\sigma^{\alpha\beta}
P_\alpha (q+q^\prime) _\beta  U(P,S)$.  
By use of the Dirac equation, one  can show  that  
1) $\bar U(P^\prime,S^\prime)\sigma^{\alpha\beta}P_\alpha 
P^\prime_\beta  U(P,S)$ 
is equivalent  to the scalar Dirac structure; and 
2) Both $\bar U(P^\prime,S^\prime)\sigma^{\alpha\beta}(q+q^\prime) 
_\alpha P^\prime_\beta  U(P,S)$ 
and $ \bar U(P^\prime,S^\prime)\sigma^{\alpha\beta}P_\alpha 
(q+q^\prime)_\beta  U(P,S)$ 
reduce  to a combination of the  the vector and scalar Dirac structures. 
We choose $\bar U(P^\prime,S^\prime)
\sigma^{\alpha\beta} P _\alpha   P^\prime_\beta 
U(P,S)$ as an independent Dirac structure,  which    can  be matched 
 with $ -(q\cdot q^\prime)g^{\mu\nu}+q^{\prime\mu} q^\nu,$  
$A^\mu B^\nu,$  $ A^\mu_1 B^\nu+ A^\mu B^\nu_1,$
 $ A^\mu_1 B^\nu- A^\mu B^\nu_1,$ and  $ A^\mu_1 B^\nu_1.$

Thus, we have specified  9 independent  chiral-odd structures for the 
non-collinear double  VCS off the massive lepton.

Now we are in a position to  make our suggestion about the 
 Lorentz decomposition of the non-collinear
 double VCS  off the massive lepton: 
{ \small
\begin{eqnarray} 
T^{\mu\nu}& = & \frac{ -(q\cdot q^\prime) g^{\mu\nu} 
+q^{\prime\mu} q^\nu  }{su} 
    \bar U(P^\prime,S^\prime) \Big(f_1 
( \rlap/q + \rlap/q^\prime)  
+f_2  \frac{ 
i \sigma^{\alpha\beta}P _\alpha   P^\prime_\beta 
}{ 2m }
     \Big)U(P,S)  \nonumber \\ 
&& +  \frac{A^\mu B^\nu}{s u }  
\bar U(P^\prime,S^\prime)\Big(f_3 ( \rlap/q + \rlap/q^\prime)   
+f_4 \frac{   
i \sigma^{\alpha\beta}P _\alpha   P^\prime_\beta 
}{ 2 m }
     \Big)U(P,S)
\nonumber \\ 
&& + \frac{A^\mu_1 B^\nu + A^\mu B^\nu_1}{s u } 
\bar U(P^\prime,S^\prime)\Big(f_5 ( \rlap/q + \rlap/q^\prime)   
+f_6 \frac{ 
i \sigma^{\alpha\beta}P _\alpha   P^\prime_\beta 
}{ 2 m } 
     \Big)U(P,S)
\nonumber \\
&& +\frac{A^\mu_1 B^\nu - A^\mu B^\nu_1}{ su }   
\bar U(P^\prime,S^\prime)\Big(f_7 ( \rlap/q + \rlap/q^\prime)   
+f_8 \frac{ 
i \sigma^{\alpha\beta}P _\alpha   P^\prime_\beta 
}{ 2 m } 
    \Big)U(P,S)
\nonumber \\ 
&& +\frac{A^\mu_1 B^\nu_1}{ s u }  
\bar U(P^\prime,S^\prime)\Big(f_9 ( \rlap/q + \rlap/q^\prime)   
+f_{10} \frac{ 
i \sigma^{\alpha\beta}P _\alpha   P^\prime_\beta 
}{  2 m } 
     \Big)U(P,S) 
     \nonumber \\ 
 &&+\frac{i  \epsilon^{\mu\nu \alpha \beta} 
q_\alpha  q^\prime_\beta}{ s u  } 
 \bar U(P^\prime,S^\prime) \Big(g_1 ( \rlap/q + \rlap/q^\prime)
\gamma_5  
+ g_2 \frac{ (P\cdot P^\prime)\gamma_5}{2m}
    \Big)U(P,S)
 \nonumber \\ 
&&+ \frac{ i( A^\mu D^\nu   + C^\mu B^\nu) }{(P\cdot P^\prime)
s u  }
\bar U(P^\prime,S^\prime) \Big(g_3 ( \rlap/q + \rlap/q^\prime)
\gamma_5  
+g_4 \frac{ (P\cdot P^\prime)\gamma_5}{2m}  
    \Big) U(P,S)
 \nonumber \\ 
&&+\frac{i(A^\mu_1 D^\nu + C^\mu B^\nu_1) } {(P\cdot P^\prime)
 s u  }
 \bar U(P^\prime,S^\prime) \Big(g_5 ( \rlap/q + \rlap/q^\prime)
\gamma_5  
+ g_6\frac{(P\cdot P^\prime)\gamma_5}{2m}  
    \Big)U(P,S)
  \nonumber \\ 
&&+\frac{i( A^\mu_1 D^\nu -C^\mu B^\nu_1)  } {(P\cdot P^\prime)
 s u }
  \bar U(P^\prime,S^\prime) \Big(g_7 ( \rlap/q + \rlap/q^\prime)
\gamma_5  
+  g_8\frac{ (P\cdot P^\prime)\gamma_5}{2m} 
  \Big) U(P,S) 
  , \label{decom} 
\end{eqnarray} 
}
where $f_i$ and $g_i$  are dimensionless  complex form factors,  
dependent  on   $s$,  $u$,    $q^2$ and  $q^{\prime 2}$.  
From our procedure to establish the above decomposition, the reader 
can convince himself  that our decomposition is complementary
in the non-collinear case. 
The $s$  and  $u$  factors in  Eq. (\ref{decom}) can be understood 
as   the remnants of the $s$- and $u$-channel propagators.

The  Compton scattering off the proton and that off the lepton 
observe  the same symmetries, while the  proton is a 
composite object.  Therefore, Eq. (\ref{decom}) applies as well to 
the  non-collinear  double  VCS  off the proton. 
 The  soft physics in the proton, in relation 
to the chiral symmetry  breaking, does not bring about  extra problems  in 
decomposing   the VCS amplitude. From now on, we understand 
 Eq. (\ref{decom}) to be the general decomposition of the double VCS tensor 
for the proton.

 By construction,  the form factors in  Eq. (\ref{decom}) are 
either symmetric or antisymmetric under the crossing and time 
reversal transformations.  The crossing symmetry properties 
of the form factors can be read off straightforwardly. 
 To obtain the time-reversal transformation properties of the form 
factors, just note that the form factors 
are functions of $s$, $u$, $q^2$ and $q^{\prime 2}$, irrelevant of the 
spin state of the proton.  Then,  one can put each of the protons in 
a specific helicity state to  show how the form factors 
transform under the  adjoint parity-time-reversal transformations. 
We summarize the crossing and 
parity-time-reversal  transformation properties of   
the form factors   in Table  2.  The  various symmetry transformation 
properties of the form factors can be employed to perform consistency 
check of theoretical calculations.
Of more interest are  the  symmetry  properties of the form factors 
under the adjoint  crossing,  parity and time-reversal transformations, 
 where  $s$ and $u$ exchange their roles. They are especially useful 
for the  study of the  single VCS, where  the on-shell condition
$q^{\prime 2}=0$  of the final photon is usually 
implicit  in the calculations.

 Now we address the reduction of Eq. (\ref{decom}) to the 
cases of the single VCS.  Regarding the single 
VCS, the on-shell condition of  the final-state photon implies 
as well the Lorentz condition of  the  final-state photon.  
Therefore, all  of its $B^\nu_1$-related  terms    become  unobservable. 
  As a consequence,  we are left with 
12 independent, observable form factors $f_{1, 2, 3, 4,5+7,
6+8},$  and $g_{1, 2, 3, 4,5+7, 6+8}$. For convenience, we 
introduce the shorthand $f_{i\pm j}=f_i \pm f_j$.  Similarly for 
the $g$-type form factors. 
 Notice that $f_{5+7,6+8}$
 and  $g_{5+7, 6+8}$  have no specific transformation 
properties under individual crossing and parity-time-reversal
 transformation. However, they are symmetric  
under the adjoint crossing-parity-time-reversal transformation. 
More concretely, they are symmetric under $s \leftrightarrow u$. 

  Notice that the single VCS tensor contains more information than 
the corresponding transition  amplitude. 
Generally speaking, $f_{5-7, 6-8, 9,10}$ and   
$g_{5-7, 6-8, 9, 10}$ do not vanish at $q^{\prime 2}=0$  on their own. 
They do not make contributions to the single VCS amplitude 
 simply because  the contraction of their associated Lorentz tensors 
with the polarization vector of the final-state photon
vanish. To obtain  these form factors, one could extrapolate  the data 
from $q^{\prime 2}\neq 0$, which is beyond the scope of this work.

Another interesting reduction of  Eq. (\ref{decom})  is  to go 
to the non-collinear  RCS.  Imposing the on-shell condition both on 
 the initial and final-state photons,  we are left with 
8 $observable$, independent form factors: $f_{1, 2, 3, 4}$ and 
$g_{1, 2, 3, 4}$. In other words, we can reach the non-collinear 
RCS by  simply dropping  those terms constructed with  $A^\mu_1$ and/or 
$B^\nu_1$.  Our conclusion is 
consistent with the independent helicity counting by 
Kroll, Sch\"urmann and Guichon \cite{kroll},
 but disagrees with  the claim made by 
Berg and Lindner that there are only 6 non-vanishing form factors 
for the   non-collinear RCS.

   Here we remark that the 
Berg-Lindner claim was incorrect, because it was
 based on an abuse of 
 the crossing symmetry of the Compton scattering. It is 
literally true that the single VCS form factors 
depend on  only 3 independent 
kinematical variables if the on-shell condition of the 
final photon is taken into account.  As far as  the 
crossing symmetry properties are concerned, however,  
 $q^{\prime 2}$ must be taken  as an independent kinematical 
variable  as  the single VCS is discussed.  In Ref. 
\cite{lindner} it is  
assumed that the single VCS form factors are three-argument  functions, 
so its  discussion about the   crossing symmetry properties are 
incorrect.  In addition, the crossing 
symmetry  for the proton was employed in Ref.\cite{lindner} to
eliminate two form factors. We  note  that the fermion  crossing 
symmetry, which is essentially a charge conjugation symmetry, 
can be used to  relate the Compton scattering off the proton to 
that off the anti-proton, so it does not generate  any constraints 
on the  VCS form factors. 

A straightforward application of Eq. (\ref{decom}) is to 
recover the manifest electromagnetic 
 gauge invariance in the leading twist 
expansion  of the DVCS tensor \cite{ji1,ra1,chen}.  Notice that in these 
leading twist expansions, all of the involved  nonperturbative matrix 
elements, such as Ji's off-forward parton distributions (OFPD) 
and  Radyushkin's  double   distributions, 
 are color gauge invariant by definition. 
In addition, all the leading twist contributions in these 
expansions  are  associated with the Lorentz structure 
of types $g^{\mu\nu} +\cdots $ and $\epsilon^{\mu\nu \cdots}$. 
 To recover the electromagnetic gauge invariance, 
one  can simply replace in Refs. \cite{ji1,ra1,chen} $g^{\mu\nu} +\cdots $
 and $\epsilon^{\mu\nu \cdots}$
by $  g^{\mu\nu} 
-q^{\prime\mu} q^\nu/(q\cdot q^\prime)$ and $\epsilon^{\mu\nu \alpha \beta} 
q_\alpha  q^\prime_\beta$ respectively, 
without need to  look into non-leading terms. 

 The usefulness of our decomposition of the non-collinear
Compton scattering tensor  is more than above. It can be 
expected  that the study of the VCS will inevitably go beyond 
leading twist and leading order.  In fact,  
some progress along this line has been witnessed \cite{os}. 
The basic use of Eq. (\ref{decom})
lies in  helping theoreticians organize complicated 
calculations  in the study of higher-twist and higher-order
contributions. 

In the following, we illustrate such procedures 
by expanding the  Born-level amplitude for the double VCS 
off the massless lepton in terms of form factors. 
Since we are working with the massless lepton,  there are only 9 
double VCS form factors. Namely,   all  the form factors 
with an even subscript drop out in Eq. (\ref{decom}) now.  
From this heuristic example,  we will verify our analysis of the 
symmetry properties of the Compton form factors. 
In addition,  we will learn that there 
do exist some physical quantities in Nature 
 that cannot be accessed directly 
by experiments  but can,   in 
principle,  be extracted by extrapolation.

To project out the form factors, we  first multiply  both sides of 
Eq. (\ref{decom}) with  the complex conjugate of 
$\bar u(p^\prime,s^\prime) 
( \rlap/q + \rlap/q^\prime)   u(p,s) $, 
and   perform the spin sum  over the initial and 
final-state  leptons so as to eliminate the Dirac  spinors. 
 Here  the   $g$-type form factors 
drop out  because the lepton has been assumed to be massless. 
Then,   we saturate  the Lorentz indices  of 
the  resulting tensor equations   in turn  with  the Lorentz 
tensors associated with each form factor.  As a result, we  
obtain  for the $f$-type  form factors 
the  algebraic equations of the following form 
\begin{equation} 
\left(
\begin{array}{c} 
c_1\\
c_3\\
c_{5+7} \\
c_{5-7} \\
c_9
\end{array} \right)= 
\left(
\begin{array}{ccccc} 
a_{11} &    a_{12} &   a_{13} &  a_{14}  &  a_{15}    \\ 
a_{21} &    a_{22} &   a_{23} &  a_{24}   &  a_{25}  \\ 
a_{31} &    a_{32} &   a_{33} &  a_{34}  &   a_{35}  \\ 
a_{41} &    a_{42} &   a_{43} &  a_{44}   &  a_{45}   \\ 
a_{51} &    a_{52} &   a_{53} &  a_{54}   &  a_{55}    \\ 
\end{array} \right)
\left(
\begin{array}{c} 
f_1\\
f_3\\
f_{5+7} \\
f_{5-7} \\
f_9 
\end{array} \right) \ . 
\end{equation}  
Similarly, by  employing   
$\bar u(p^\prime,s^\prime) 
( \rlap/q + \rlap/q^\prime) \gamma_5   u(p,s) $, 
 we have for  the $g$-type form factors, 
\begin{equation} 
\left(
\begin{array}{c} 
d_1\\
d_3\\
d_{5+7} \\
d_{5-7} 
\end{array} \right)=\left(
\begin{array}{ccccc} 
b_{11} &    b_{12} &   b_{13} &  b_{14}      \\ 
b_{21} &    b_{22} &   b_{23} &  b_{24}     \\ 
b_{31} &    b_{32} &   b_{33} &  b_{34}    \\ 
b_{41} &    b_{42} &   b_{43} &  b_{44}   
\end{array} \right) 
\left(
\begin{array}{c} 
g_1\\
g_3\\
g_{5+7} \\
g_{5-7} 
\end{array} \right) \ , \end{equation}  
To save space, we omit the concrete expressions for 
 $c_i$, $d_i$,  $a_{ij}$ and $b_{ij}$. 
 
 Straightforward algebra  gives us 
\begin{eqnarray} 
f^{(0)}_1(s,u,q^2,q^{\prime 2})&=& \frac{s - u}{s+u}\ ,\label{lept10}   \\ 
f^{(0)}_3(s,u,q^2,q^{\prime 2})
&=& \frac{(s-q^2) (u-q^2) (s-q^{\prime 2} ) (u-q^{\prime 2} )} 
{(s^2 - u^2)  (su- q^2q^{\prime 2})}\ , \\   
f^{(0)}_{5+7}(s,u,q^2,q^{\prime 2})
 &=& -\frac{ (s-q^{\prime 2} ) (u-q^{\prime 2} )
(s + u -q^2) } { (s-u)(su- q^2q^{\prime 2})}\ , \\  
 f^{(0)}_{5-7}(s,u,q^2,q^{\prime 2})
 &=& -\frac{ (s-q^{\prime 2} ) (u-q^{\prime 2} )
(s + u -q^{\prime 2}) } { (s-u)(su- q^2q^{\prime 2})}\ , \\  
 f^{(0)}_9(s,u,q^2,q^{\prime 2})
&=&\frac{ (s+u) [s^2 +u^2 + su  -(q^2 +q^{\prime 2} ) 
( s+u) + q^2q^{\prime 2}]} { (s-u)(su- q^2q^{\prime 2})}\ , \\ 
 g^{(0)}_1(s,u,q^2,q^{\prime 2})&=& \frac{ (q^2 -q^{\prime 2} )
(s+u)[s^2 +u^2 - (q^2 +q^{\prime 2} ) 
( s+u) +2 q^2q^{\prime 2}]} 
{ [2 s u-  (q^2 +q^{\prime 2} ) ( s+u)+ q^4 + q^{\prime 4}] 
[s^2 +u^2 + 2 su - 4  q^2q^{\prime 2}]} \ , \\ 
g^{(0)}_3(s,u,q^2,q^{\prime 2})
&=&\frac{ (s-q^{\prime 2} ) (u-q^{\prime 2} )(s-q^2) (u-q^2)
(s + u -q^2 -q^{\prime 2}) } 
{(s-u) (su - q^2 q^{\prime 2}) 
[2 s u-  (q^2 +q^{\prime 2} ) ( s+u)+ q^4 + q^{\prime 4}]}\ , \\ 
g^{(0)}_{5+7}(s,u,q^2,q^{\prime 2})
 &=&\Big\{ (s + u)(s + u -q^2 -q^{\prime 2})
[s^2 +u^2 - (q^2 +q^{\prime 2} ) ( s+u) + q^2q^{\prime 2}]\nonumber \\ 
& &  ~~~~~ 
[-su (s +u)  +q^{\prime 2} (s^2 +u^2) +q^2q^{\prime 2}(s +u) 
+ 2  q^2q^{\prime 4} ] \Big\}\nonumber \\
& & \times  
\Big\{ (s-u)(s u - q^2q^{\prime 2})[s^2 +u^2 + 2 su - 4  q^2q^{\prime 2}]   
\nonumber \\  && ~~~~~ 
[2 su -  (q^2 +q^{\prime 2} ) ( s+u)+ q^4 + q^{\prime 4}] \Big\} ^{-1} \ , \\ 
g^{(0)}_{5-7}(s,u,q^2,q^{\prime 2}) &=&\Big\{ (s + u)(s + u -q^2-q^{\prime 2})
[s^2 +u^2 - (q^2 +q^{\prime 2} ) ( s+u) + q^2q^{\prime 2}]\nonumber \\
& &~~~~~ [-su(s +u) +q^{ 2} (s^2 +u^2) +q^2q^{\prime 2}(s +u) 
+ 2  q^4q^{\prime 2} ] \Big\}\nonumber \\
 & & \times 
 \Big\{(s-u) (s u - q^2q^{\prime 2}) [s^2 +u^2 + 2 su - 4  q^2q^{\prime 2}] 
\nonumber \\
&& ~~~~~   
[2 su -  (q^2 +q^{\prime 2} ) ( s+u)+ q^4 + q^{\prime 4}]
\Big\} ^{-1} \ , \label{lept20} 
\end{eqnarray} 
where the  superscript $(0)$  labels the lowest-order results.
Obviously, Eqs. (\ref{lept10}-\ref{lept20})  are 
consistent with our general symmetry analysis summarized in Table. 2. 

Letting $q^{\prime 2}=0$ in Eqs. (\ref{lept10}-\ref{lept20}),
we obtain the form factors for the single VCS off the massless lepton. 
Evidently, in the single VCS case, $f_{5-7}$,  $f_9$ and 
$g_{5-7}$  do not vanish  themselves. 
Moreover, all the form factors have specific symmetry properties 
under $s\leftrightarrow u$, while $f_{5-7}$ and $g_{5-7}$ have no 
specific symmetry properties under $q^2 \leftrightarrow q^{\prime 2}$.

As  we  have  pointed out,  $f_{5-7}$, 
 $f_9$ and $g_{5-7}$  do not come into 
action in the  single VCS amplitude,  
due to the Lorentz conditions of the 
final photon.  Such a fact informs us that they can be
replaced by  arbitrary numbers,
in  any  complete,   gauge-invariant expansions of the 
single VCS tensor for the massless  lepton (quark). 
In another word, even   $f_{5-7}$, $f_9$ and $g_{5-7}$  
are included explicitly in  the bases  for expanding the 
single VCS tensor,  they   cannot  be solved  out uniquely.
 By  explicit calculations, 
one can easily show that  all $a_{ij}$ and $b_{ij}$ with  
 index 4 and/or 5  vanish  in the case of the single VCS.  
Hence,   $c_{5-7} $,   $c_9$ and $d_{5-7} $ must 
vanish, which is a manifestation of the 
electromagnetic gauge invariance. 
The above fact sounds trivial to the expansion of the Born 
amplitudes,  but  serves as a very useful 
consistency check in the practical calculations of loop corrections. 
Note that one can also construct projectors (with Lorentz tensor 
 and Dirac bilinear structures) for all of the form factors that 
function in the single VCS.

 Our  form factor decomposition of the non-collinear Compton 
scattering tensor has further implications to the study 
of Ji's OFPDs and Radyushkin's double distributions.
In the leading-twist Feynman diagram expansion of 
  the DVCS off the proton,  the 
underlying dynamics is believed to be the single 
VCS off the massless quark. A virtue of our decomposition, 
Eq. (\ref{decom}), is that  the Lorenz tensors for single VCS tensor 
of the massless quark are the same as those in the DVCS tensor of the 
proton. Since the VCS off the quark and that off the proton are subject 
to  the same symmetry constraints, one can naturally conclude that 
both the  OFPDs and double distributions 
 possess some symmetry properties. 
These symmetry properties will impose some  constraints 
as one attempts to model  the OFPDs and double distributions. 

In Ji's expansion \cite{ji1} of the  DVCS tensor,  a light-like momentum 
$p^\mu$  in connection with the  average   of the initial- 
and final-state proton momenta is introduced. Then,  the momenta of the 
initial- and final-state protons are approximated as $(1+\xi )p$ 
and  $(1-\xi )p$, respectively, where $\xi$ ($0<\xi <1$) 
is the analog of the 
Bjorken variable. Correspondingly, the  momenta of 
the initial- and final-state partons participating in  the hard 
single VCS  are effectively taken as  $(x+\xi )p$ 
and  $(x-\xi )p$.  One can easily show that two partonic Mandelstam 
variables are related to  their hadronic counterparts via 
\begin{eqnarray} 
\hat s & \equiv &  [q^\prime +(x-\xi )p]^2 \simeq \frac{x-\xi}{1-\xi} s \ , \\ 
\hat u & \equiv &  [q^\prime -(x+\xi )p]^2\simeq \frac{x+\xi}{1+\xi} u \ . 
\end{eqnarray} 
A DVCS form factor of the proton can be roughly 
thought of as the convolution of the corresponding parton 
form factor with an  appropriate  OFPD. The quark form factor is 
either symmetric or antisymmetric under   
$\hat s \leftrightarrow \hat u$, which  amounts to $s\leftrightarrow u$ and 
$\xi \to -\xi$.  
Hence,   the symmetry properties of the proton form factor demands  that 
the   OFPDs  satisfy the following relations: 
\begin{eqnarray} 
H(x, \xi, \Delta^2)&= &H(x, -\xi, \Delta^2)\ , \label{01}  \\ 
E(x, \xi, \Delta^2)&= & E(x, -\xi, \Delta^2)\ ,  \\ 
\tilde H(x, \xi, \Delta^2)&= &\tilde H(x, -\xi, \Delta^2)\ ,  \\ 
\tilde E(x, \xi, \Delta^2)&= &\tilde E(x, -\xi, \Delta^2)\ , \label{04}  
\end{eqnarray}
where $\Delta^2\equiv (P^\prime -P)^2$ is  the Mandelstam variable $t$. 
In fact, the above relations can be derived 
from  the definitions of these OFPDs directly by   time reversal invariance.  
Further, one can show \cite{ji}
 that  all of the  OFPDs are real, with the help of 
Eqs. (\ref{01}-\ref{04}). 

 Now  let us consider Radyushkin's expansion \cite{ra1} of the DVCS 
tensor, in which  the  momenta of the initial- and final-state partons
are approximated  as  $xp +yr$ and $xp -\bar y r$ respectively, with 
$r\equiv P-P^\prime$ and $\bar y= 1-y$.  Here two partonic Mandelstam variables read 
\begin{eqnarray} 
\hat s & \equiv &  [q^\prime +xp-(1-y)r]^2 \simeq \bar y (s+u) -  x u \ , \\ 
\hat u & \equiv &  [q^\prime -xp-yr]^2\simeq  -y (s +u) - x u \  .   
\end{eqnarray} 
 Obviously, $\hat s \leftrightarrow \hat u$ implies 
$y\to - \bar y$ and $\bar y \to -y$. Now,  the nonperturbative physics 
is incorporated by two double distributions $F(x,y)$ and $G(x,y)$. 
 Consequently,  $F(x,y)$ and $G(x,y)$ must be invariant 
under  the transformation
$y\to - \bar y$ and $\bar y \to -y$.  Here we recall that 
 $F(x,y)$ is actually  defined by the following leading-twist
expansion of the proton  matrix: 
\begin{equation}  
\int \frac{d \lambda d \eta} 
{(2\pi)^2} e^{i \lambda (x+\zeta  y) -i \eta  (x-\bar y \zeta)} 
 \langle P^\prime,S^\prime| \bar \psi (\lambda n) 
\gamma^\alpha \psi(\eta n)|P, S\rangle  = \bar U(P^\prime,S^\prime) 
 \gamma^\alpha U(P,S)  F(x, y) +\cdots , \label{Fxy} 
 \end{equation}  
where $\zeta\equiv r\cdot n$ and 
  $ n$ is a light-like vector with an inverse momentum dimension.
Hence, we can effectively write down 
\begin{eqnarray} 
F(x,y)\equiv F(x;y,\bar y), 
\end{eqnarray}
That is to say, $y$ and $\bar y$ function  in the 
double distributions  as if they were  two independent 
variables. To examine the symmetry properties of $F(x,y)$, 
one can put each of  the 
protons in a helicity  eigenstate.  Then, by use of  time reversal 
invariance,  one can quickly show
\begin{eqnarray}  
F(x;y,\bar y)&=&F(x; -\bar y \ , - y )  \ ,  \label{sy1}
\end{eqnarray}   
 Similarly, there is 
\begin{eqnarray}  
G(x;y,\bar y)&=&G(x; -\bar y \ , - y )  \ .  \label{sy2}
\end{eqnarray}   
Equations (\ref{sy1}-\ref{sy2}) are a useful 
guide as one parameterizes  $F(x,y)$ and $G(x,y)$. 

In  Ref. \cite{ra1},  there was an observation that  the double 
distributions are  purely real  in some toy models. In fact,  this 
is generally true  in QCD. To show this,  just  take the complex conjugate of 
Eq. (\ref{Fxy}).  There will be  $F^\ast(x;y,\bar y)=F(x; -\bar y \ , - y )$.
Combining  this with Eq. (\ref{sy1}), we know that $F(x,y)$ is real.  
The proof that  $G(x,y)$ is real goes the same way.

  In closing, we remark  the limitations of our form factor 
description of the Compton scattering tensor. 
It  is applicable to the non-collinear Compton scattering, both real and virtual.
 As going to the collinear limits, however,  it becomes pathological. 
The case   is  the worst 
 as one attempts to discuss the transverse proton spin 
dependence of the Compton amplitude in the collinear limits. There is 
no remedy in  our present  scheme to  parameterize the Compton scattering 
tensor in terms of form factors.  In fact,  the drawbacks of our 
decomposition are shared by all of the  present  Feynman diagram expansions 
and OPE analyses of the proton DVCS tensor.  To develop a form factor 
description of the Compton scattering tensor suitable for taking 
the collinear limits,    
one can demand  that the gamma matrices carry free Lorentz indices. 
At present, we have not seen any advantages in adopting such a scenario.

\acknowledgements 
The author  thanks  M. Anselmino, M. Diehl, M.   Gl\"uck, 
 T. Gousset, P.A.M. Guichon, Xiangdong Ji, 
Boqiang Ma,  B. Pire, A. Radyushkin, 
 J.P. Ralston  and E. Reya for useful discussions and/or correspondence.
In particular, he is grateful to M. Diehl and B. Pire for 
helping clarify the constraints of the crossing symmetry 
on  the Compton scattering tensor.

\vspace*{2cm}
Table 1. Surviving independent   helicity amplitudes of the 
various Compton scattering in the collinear scattering limits. 

\vspace*{1cm}
{\small
\begin{center} 
\begin{tabular}{c|c|c|c|c|c} 
\hline 
\multicolumn{2}{c} {$\gamma ^\ast (q)  +N(P)
 \to \gamma^\ast (q^\prime)  + N(P^\prime) $}\vline & 
\multicolumn{2}{c} {$\gamma ^\ast(q) +N(P) \to 
\gamma (q^\prime) + N(P^\prime) $}\vline   &
\multicolumn{2}{c} {$\gamma(q)  +N (P)  \to 
\gamma (q^\prime) + N (P^\prime) $}  \\
\hline 
forward& backward&forward& backward& forward& backward\\
\hline 
$A(1, \frac{1}{2};1, \frac{1}{2})$ & 
 $A(1, \frac{1}{2};-1, -\frac{1}{2})$ & 
$A(1, \frac{1}{2};1, \frac{1}{2})$ & 
$A(1, \frac{1}{2};-1, -\frac{1}{2})$ & 
$A(1, \frac{1}{2};1, \frac{1}{2})$  & 
 $A(1, \frac{1}{2};-1, -\frac{1}{2})$ \\ 
$A(0, \frac{1}{2};0, \frac{1}{2})$ & 
$A(0, \frac{1}{2};0, -\frac{1}{2})$ & 
~  & 
~  & 
~ & 
~ \\
$A(0, \frac{1}{2};-1, -\frac{1}{2})$ & 
$A(0, \frac{1}{2};1, \frac{1}{2})$ & 
 $A(0, \frac{1}{2};-1, -\frac{1}{2})$ & 
$A(0, \frac{1}{2};1, \frac{1}{2})$ & 
~ & 
 \\
$A(-1, \frac{1}{2};-1, \frac{1}{2})$ & 
$A(-1, \frac{1}{2};1, -\frac{1}{2})$ & 
$A(-1, \frac{1}{2};-1, \frac{1}{2})$  & 
 $A(-1, \frac{1}{2};1, -\frac{1}{2})$ & 
$A(-1, \frac{1}{2};-1, \frac{1}{2})$ & 
$A(-1, \frac{1}{2};1, -\frac{1}{2})$\\ \hline
\end{tabular} 
\end{center} 
}

\vspace*{2cm}

Table 2. Crossing and parity-time-reversal transformation 
properties of the  double VCS form factors. The plus  and minus 
signs represent that the form factor is symmetric 
and  antisymmetric, respectively. 

\vspace*{1cm}

\begin{center}
\begin{tabular}{c|c|c|c|c|c|c|c|c|c|c|c|c|c|c|c|c|c|c}\hline  
form factor & $f_1$ & $f_2$ &  $f_3$ & $f_4$ & 
$f_5$ & $f_6$ &$f_7$ & $f_8$ &$f_9$ & $f_{10}$
 & $g_1$ & $g_2$ &  $g_3$ & $g_4$ & 
$g_5$ & $g_6$ &$g_7$ & $g_8$
 \\
\hline  
$s \leftrightarrow   u $ \&  $ q^2\leftrightarrow    q^{\prime 2}$ 
 & $-$ & $+$  &$-$ & $+$ &$-$ & $+$ & + &$ -$ &$ -$& $+$ 
& $-$& $+$  &$ -$& $+$  &$ -$& + & +&$ -$ \\  
\hline 
 $q^2  \leftrightarrow q^{\prime 2} $
 & $+$  &$ +$ & $+$ &$ +$& $+$ & $+$& $-$ & $-$& $+$ &$ +$
& $-$& $+$& $+$ &$ -$& $+$  & $-$& $-$& $+$\\
\hline  
$s   \leftrightarrow u $
 & $-$ & $+$ & $-$ &$ +$&$ -$ & $+$& $-$ & $+$& $-$&$ +$
& $+$& $+$ & $-$ &$ -$&$ -$& $-$& $-$& $-$\\
\hline  
\end{tabular}
\end{center}

\end{document}